\documentclass[runningheads]{llncs}

\usepackage{eccv}

\usepackage{eccvabbrv}

\usepackage{graphicx}
\usepackage{booktabs}
\usepackage[utf8]{inputenc}
\usepackage[T1]{fontenc}
\usepackage{microtype}
\usepackage{xspace}
\usepackage{amsfonts}
\usepackage{amsmath}
\usepackage{amssymb}
\usepackage{mathtools}
\usepackage{nicefrac}
\usepackage{xcolor}
\usepackage{multirow}
\usepackage{array}
\usepackage{enumitem}
\usepackage{placeins}
\usepackage{url}
\usepackage{microtype}
\usepackage{float}
\usepackage{seqsplit}
\newcommand{\tgvlm}{\textsc{TractoGraphVLM}\xspace}

\usepackage[accsupp]{axessibility}  

\usepackage{hyperref}

\usepackage{orcidlink}

\begin{document}

\title{TractoGraphVLM: A Unified Vision-Language Framework for White Matter Tractography}
\titlerunning{TractoGraphVLM}

\author{Gurucharan Marthi Krishna Kumar\inst{1}\orcidlink{0000-0002-8069-6508} \and
Janine Dale Mendola\inst{2}\orcidlink{0000-0001-8286-2517} \and
Amir Shmuel\inst{3}\orcidlink{0000-0003-3028-6639}}

\authorrunning{G. Marthi Krishna Kumar et al.}

\institute{Montreal Neurological Institute, McGill University, Montreal, Canada\\
\email{gurucharan.marthikrishnakumar@mail.mcgill.ca} \and
Department of Ophthalmology, McGill University, Montreal, Canada\\
\email{janine.mendola@mcgill.ca} \and
McConnell Brain Imaging Centre, Montreal Neurological Institute, Departments of Neurology, Neurosurgery, Physiology and Biomedical Eng., McGill University, Montreal, QC, Canada\\
\email{amir.shmuel@mcgill.ca}}

\maketitle

\raggedbottom

\begin{abstract}
Vision language models have transformed two dimensional medical imaging, yet extending them to three dimensional white matter tractography remains challenging due to the complex topology of continuous fiber bundles. We introduce \tgvlm, a unified framework for four tasks: white matter bundle classification, text to tract retrieval, anatomical captioning, and visual question answering. It is built on a shared GPS graph architecture, training procedure, and read-out design. Fiber bundles are represented as streamline graphs whose nodes encode 3D position and tangent orientation. A General, Powerful, Scalable (GPS) graph transformer produces bundle embeddings aligned with a frozen BiomedBERT text encoder via contrastive learning, while a BioGPT decoder with visual prefix tokens generates captions and answers. A single model, with one shared encoder and decoder, is trained jointly across all four tasks and evaluated from one checkpoint. Trained on $1{,}113$ HCP Young Adult subjects, \tgvlm achieves $91.8\%$ white matter bundle classification accuracy, $84.7\%$ retrieval R@1, BLEU-4=20.1, ROUGE-L=66.8, and $66.4\%$ VQA accuracy on a held out test set. The same checkpoints transfer zero-shot to $725$ HCP Aging subjects with a modest drop on the discriminative tasks and a larger but still coherent one on the generative tasks, indicating robustness to age and acquisition shift. Language supervision also yields richer representations than label-only training, recovering anatomical structure such as hemisphere and fiber family that is carried by the captions but never supplied as an explicit classification label. Swapping only the visual encoder, we find that graphs preserving fiber orientation outperform volumetric baselines, with GPS giving the best overall balance. Our generative metrics measure consistency with a structured knowledge base rather than independent clinical text; even so, \tgvlm shows that classifying, retrieving, describing, and answering questions about a white matter bundle can be served by a single jointly trained model that learns transferable neuroanatomy from language alone. Our codebase, including all training and evaluation pipelines, is publicly available at \url{https://github.com/AS-Lab/Marthi-et-al-2026-TractoGraphVLM-Unified-Vision-Language-White-Matter-Tractography}
\end{abstract}

\section{Introduction}
\label{sec:intro}

White matter tractography reconstructs brain connectivity non-invasively from diffusion MRI~\cite{basser1994,jeurissen2019diffusion} and is increasingly investigated for applications in clinical neuroscience. Fiber bundle abnormalities are implicated in multiple sclerosis, stroke, and neurodegenerative diseases~\cite{ciccarelli2008}, while tractography increasingly informs pre-surgical planning~\cite{essayed2017}. Despite its importance, tractography analysis remains a purely structural pipeline: complex 3D bundles are typically interpreted through manual visual inspection, a process that does not scale to large population cohorts.

Recent advances in vision-language models (VLMs) have transformed two-dimensional medical imaging by coupling visual understanding with natural-language explanation~\cite{bannur2023,lu2024,yan2025multimodal}. Extending this paradigm to tractography, however, faces a unique challenge: a fiber bundle is not a simple image but a collection of thousands of continuous streamlines whose fine directional structure is poorly captured by voxel grids. This raises the core question of this work: \emph{how should white matter fiber bundles be represented for effective vision-language understanding?}

Two main representation paradigms exist. \textbf{Volumetric} approaches rasterize streamlines into occupancy grids for 3D CNNs~\cite{tran2015,hara2018can} or Vision Transformers~\cite{dosovitskiy2020image,hatamizadeh2022unetr}. While convenient for reusing image-based architectures, they discard sub-millimetre directional detail. \textbf{Graph-based} approaches model streamlines as graphs with nodes carrying 3D position and orientation~\cite{kipf2016gcn,velivckovic2017gat}, preserving continuous topology. No prior work has systematically compared these paradigms inside a unified multi-task VLM framework, and we close this gap with a controlled comparison in which only the visual backbone varies. Our central finding is that the choice of encoding, more than task-specific heads, determines how effectively a bundle can be named, described, and queried in natural language.

\paragraph{Related work.}
Deep-learning methods for tractography have largely remained single-task. TractCloud~\cite{xue2023tractcloud} uses point-cloud encoding for whole-brain segmentation; FINTA~\cite{legarreta2021finta} and FIESTA-AE~\cite{dumais2023fiesta} apply autoencoders for streamline filtering; PointNet-style models~\cite{qi2017pointnet,qi2017pointnet2} support bundle classification~\cite{gupta2017}. Even where these methods address more than one objective, they stay within a discriminative output space of labels, masks, or reconstructions, and none align bundle geometry with a shared text embedding. All produce categorical outputs and lack the ability to generate text, retrieve bundles via natural-language queries, or answer open-ended questions. Meanwhile, general medical VLMs~\cite{zhang2022,wang2022medclip,li2023llavamed,zhang2023biomedclip} operate on 2D image-text pairs and have not been extended to 3D streamline data. The two lines of work are thus complementary but disjoint: language-capable models do not handle streamlines, and streamline-capable models do not handle language. \tgvlm bridges them by bringing a multi-task vision-language interface to a topology-preserving graph representation of tractography.

\paragraph{Contributions.}
\begin{enumerate}[leftmargin=*,itemsep=2pt]
  \item \textbf{Unified tractography VLM.} \tgvlm is a single model with a shared encoder and decoder that jointly performs bundle classification, text-to-tract retrieval, anatomical captioning, and visual question answering (VQA).
  \item \textbf{Representation study.} Through controlled comparisons that substitute the visual encoder while keeping the rest of the framework fixed, we find the General, Powerful, Scalable (GPS) graph transformer to be the most effective overall backbone in our comparison, particularly on generative tasks, while remaining competitive on discriminative metrics.
  \item \textbf{Value of multi-task language supervision.} Linear probing shows that VLM training produces consistently richer representations than supervised classification, recovering anatomical structure (hemisphere and broad fiber family) more accurately; although this structure is never supplied as an explicit classification label, it is present in the caption text the VLM is trained on.
  \item \textbf{Cross-dataset generalization.} HCP Young Adult-trained checkpoints transfer zero-shot to the HCP Aging cohort across all four tasks, indicating robustness to age and acquisition shift rather than dependence on a single acquisition protocol.
\end{enumerate}

\section{Methods}
\label{sec:methods}

\begin{figure}[!ht]
  \centering
  \includegraphics[width=\linewidth]{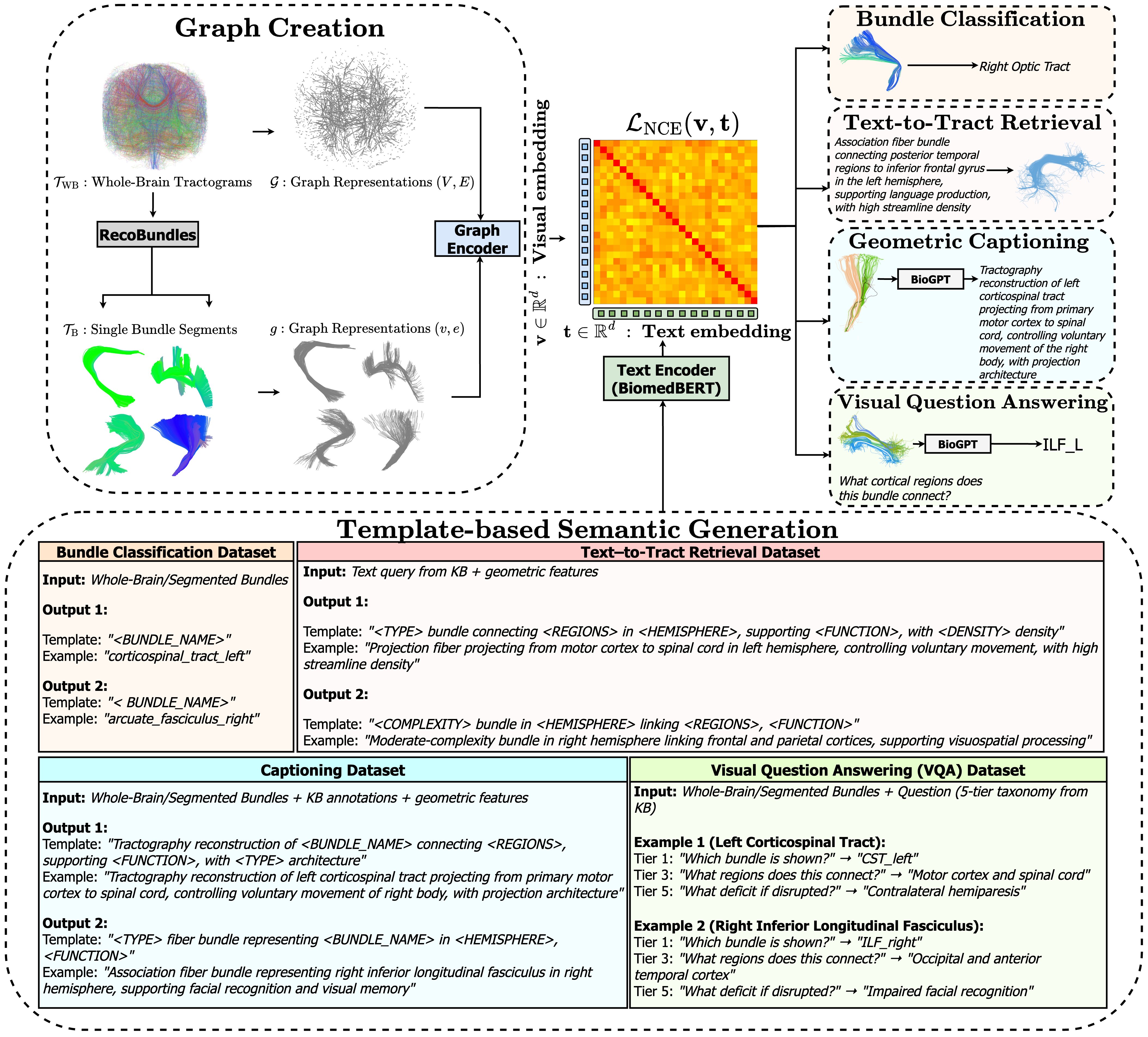}
  \caption{\textbf{\tgvlm framework.} Whole-brain tractograms are segmented into bundles (RecoBundles) and encoded as streamline graphs, which a GPS encoder maps to a visual embedding~$\mathbf{v}$ (left). A symmetric InfoNCE loss~$\mathcal{L}_{\mathrm{NCE}}(\mathbf{v},\mathbf{t})$ aligns $\mathbf{v}$ with a frozen BiomedBERT text embedding~$\mathbf{t}$; the captions, queries, and questions come from templates grounded in a neuroanatomical knowledge base (bottom). Four read-outs share this embedding (right): classification by nearest prototype, retrieval by ranking bundles against a query, and captioning and VQA by feeding~$\mathbf{v}$ to a BioGPT decoder as visual prefix tokens.}
  \label{fig:framework}
\end{figure}

\tgvlm is a unified vision-language framework for white matter tractography with two main stages. First, each fiber bundle is converted into a geometric graph whose nodes encode 3D position and local orientation (Figure~\ref{fig:s2g}). Second, a graph encoder produces a visual embedding from these graphs that is aligned with text via contrastive learning, while a decoder generates anatomical descriptions and answers visual questions (Figure~\ref{fig:framework}). The model is trained jointly in a single run: one shared GPS encoder and one BioGPT decoder are optimised across all four tasks under a combined objective, following a two-stage curriculum detailed in Section~\ref{sec:arch}. At evaluation, a single checkpoint serves every task, with each read-out head predicting from the same shared embedding.

\subsection{Datasets}
\label{sec:dataset}

\textbf{HCP Young Adult (HCP-YA).} We use multi-shell diffusion MRI from $1{,}113$ subjects~\cite{van2013wu} ($b = 1000, 2000, 3000\,\mathrm{s/mm^2}$, $90$ directions per shell, $1.25\,\mathrm{mm}$ isotropic resolution). Whole-brain tractograms are generated with iFOD2~\cite{tournier2010} under anatomically constrained tractography~\cite{smith2012} and segmented into the $78$ bundles of the HCP-842 atlas~\cite{yeh2018,bullock2022} using RecoBundles~\cite{garyfallidis2018}. Subjects are partitioned $70/15/15\%$ into training, validation, and test sets with no subject overlap. All model development uses only this cohort.

\textbf{HCP Aging.} We evaluate zero-shot generalization on $725$ subjects from the HCP Aging cohort~\cite{bookheimer2019}, acquired at $1.5\,\mathrm{mm}$ isotropic resolution under a different protocol. The identical processing pipeline (iFOD2 + anatomically constrained tractography + RecoBundles segmentation) is applied, yielding the same $78$ bundles and graph representation. HCP Aging is never seen during training.

\textbf{Anatomical knowledge base.} We compile a structured knowledge base for all $78$ bundles from standard neuroanatomical references~\cite{catani2012atlas,schmahmann2008,wakana2007} and clinical/topographical reviews~\cite{essayed2017,radwan2022}. Each entry includes trajectory, endpoints, functional roles, lateralisation, length range, and clinical associations (representative entries in the supplementary material, Sec.~\ref{app:kb}).

\subsection{Streamline-to-Graph Construction}
\label{sec:graph}

\begin{figure}[h]
  \centering
  \includegraphics[width=\linewidth]{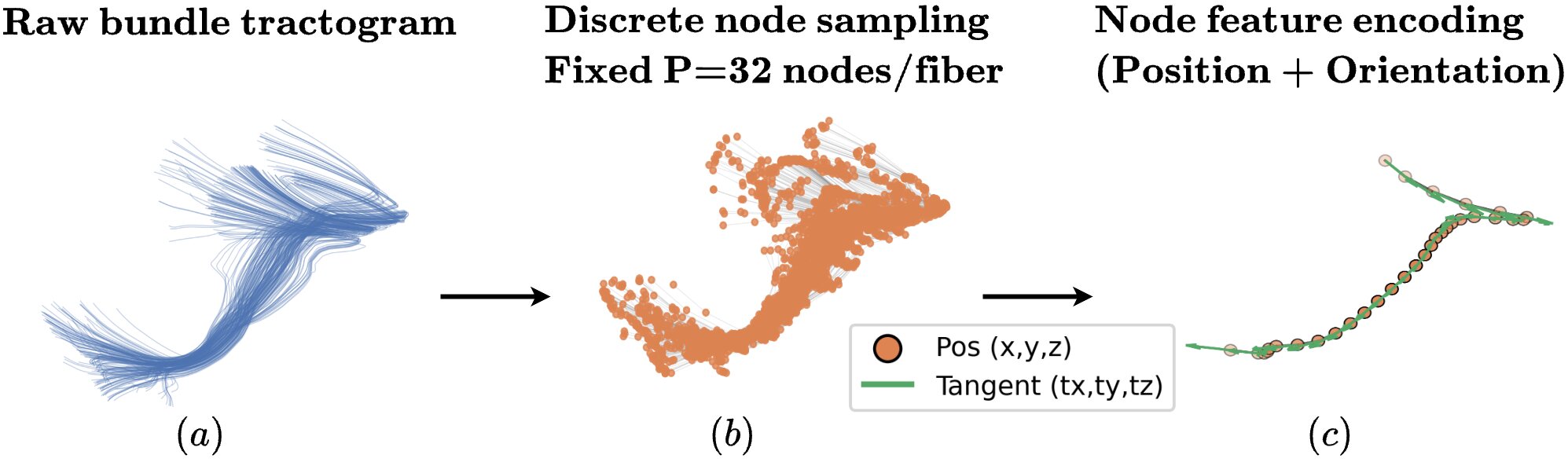}
  \caption{\textbf{Streamline-to-graph construction.} A fiber bundle (exemplified by the arcuate fasciculus) is transformed by sampling $P{=}32$ equidistant nodes per streamline. Each node carries a 6-D feature vector of 3D position and unit tangent orientation. See Section~\ref{sec:graph} for details.}
  \label{fig:s2g}
\end{figure}

Each bundle is converted into a graph $\mathcal{G} = (V, E)$ as shown in Figure~\ref{fig:s2g}. Streamlines are resampled to a fixed $P=32$ equidistant nodes (ablated in Section~\ref{sec:ablation_density}), producing a uniform node set independent of bundle length.

Each node is represented by a 6-dimensional feature vector $\mathbf{x}_i = [\mathbf{p}_i; \mathbf{t}_i]$, concatenating spatial position with the local unit tangent (computed via central finite differences). Tangent vectors are critical for disambiguating bundles with locally similar geometry (Section~\ref{sec:ablation_features}). Edges connect consecutive nodes along each streamline, capturing the sequential trajectory of individual fibers, and spatially proximate nodes across streamlines, capturing the coherence of the bundle as a whole. Laplacian positional encodings~\cite{dwivedi2023benchmark} are added to supply global structural context that local connectivity alone cannot express. Full graph construction details appear in the supplementary material, Sec.~\ref{app:arch_details}.

\subsection{Text Generation and Language Supervision}
\label{sec:textgen}

Manually annotated tractography-language pairs are impractical at scale. We therefore generate supervision programmatically from geometric features $\phi(\mathcal{T})$ and the curated knowledge base. Because the templates are populated from knowledge-base fields keyed on bundle identity, a bundle's caption, query, and answer content is largely determined by its label. We therefore treat the generative metrics as internal-consistency measures and compare against a classifier-plus-lookup control (Sec.~\ref{sec:control}) that exploits this determinacy.

\paragraph{Stage 1: Feature-grounded template instantiation.} For each bundle $\mathcal{T}$ we extract four geometric descriptors: streamline count $N$, spatial extent $E$ (bounding-box diagonal), normalized density $\rho = N/E^3$, and complexity class $\kappa \in$ \{\texttt{simple},\, \texttt{moderate},\, \texttt{complex}\} (details in the supplementary material, Sec.~\ref{app:textgen_full}). These are injected into structured templates to produce 15 caption variants and 5 VQA pairs per sample via stochastic grammar and synonym augmentation. All factual content is strictly constrained by the knowledge base.

\paragraph{Stage 2: LLM paraphrasing.} Template outputs are stylistically repetitive. Following practices established in large-scale vision--language dataset construction~\cite{li2023blip2,liu2023visual,chen2024sharegpt4v,tu2024towards}, we apply a Claude rewriting pass that diversifies surface form while preserving the anatomical claims and qualitative descriptors exactly (paraphrasing procedure detailed in the supplementary material, Sec.~\ref{app:textgen_full}). This yields up to 45 caption candidates per bundle, filtered to 15 by a ROUGE-L diversity criterion~\cite{lin2004rouge}. Unlike general-purpose captioners such as BLIP-2~\cite{li2023blip2} or InstructBLIP~\cite{dai2023instructblip}, which generate descriptions directly from pixels, our LLM rewrites strings whose factual content is already fixed by the template slots and knowledge-base lookups. This limits the LLM's freedom to introduce anatomical claims outside the source references, which is a known failure mode of unconstrained medical VQA systems~\cite{bannur2023,chen2024vision}. It does not, by itself, verify those source references.

\paragraph{VQA and retrieval.} VQA pairs span five tiers of reasoning complexity (identification $\to$ clinical function), following the hierarchical design of PathVQA~\cite{he2020pathvqa} and VQA-RAD~\cite{lau2018dataset} (templates in the supplementary material, Sec.~\ref{app:textgen_full}). Retrieval queries range from simple identity strings to compositional queries combining anatomy, function, and geometry, motivated by compositional retrieval benchmarks~\cite{baldrati2023zero,ma2023crepe}.

\subsection{Model Architecture}
\label{sec:arch}

\tgvlm consists of three components: a visual graph encoder $f_v$, a frozen text encoder $f_t$ (BiomedBERT-base)~\cite{gu2021biomedbert}, and an autoregressive decoder $g$ (BioGPT)~\cite{luo2022biogpt}. The default visual encoder is the GPS graph transformer~\cite{rampavsek2022gps}, which maps each bundle graph to a 256-dimensional embedding. For the encoder comparison (Section~\ref{sec:encoder_comparison}), we substitute alternative graph, point-cloud, and volumetric backbones (listed in the supplementary material, Sec.~\ref{app:encoders}) while keeping the remainder of the framework fixed.

To let the decoder generate text from a bundle, we project its embedding into eight ``visual prefix'' tokens and prepend them to the BioGPT input, so the decoder is conditioned on the same embedding the contrastive loss aligns with text. This sharing is deliberate: the contrastive term pulls matching bundle and text embeddings together while the generative term forces the shared embedding to retain enough detail to reconstruct a full description, so a single encoder learns a representation that is at once discriminative and descriptive. Classification and retrieval contribute only the contrastive term; captioning and VQA add the generative term. Full architectural details and parameter counts are provided in the supplementary material, Sec.~\ref{app:arch_details}. All four read-out heads consume the same visual embedding produced by the single shared GPS encoder; extending the framework to a new task reuses this encoder and requires only an additional read-out head in the joint objective.

\textbf{Training.} We optimize with AdamW~\cite{loshchilov2017decoupled} (peak LR $3{\times}10^{-4}$, weight decay $10^{-2}$, batch size 128). The total loss is
\begin{equation}
  \mathcal{L} = \mathcal{L}_{\mathrm{CLIP}} + \alpha\,\mathcal{L}_{\mathrm{LM}} + \beta\,\mathcal{L}_{\mathrm{proto}},
  \label{eq:loss}
\end{equation}
where $\mathcal{L}_{\mathrm{CLIP}}$ is symmetric InfoNCE~\cite{oord2018} (temperature $\tau$ ablated in Section~\ref{sec:ablation_temp}), $\mathcal{L}_{\mathrm{LM}}$ is autoregressive cross-entropy on captions and VQA targets ($\alpha=0.5$), and $\mathcal{L}_{\mathrm{proto}}$ is the prototype cross-entropy term used for classification ($\beta=1$). The generative and prototype terms are activated only in the second training stage; in the first stage $\alpha=\beta=0$. Training uses a two-stage curriculum: 250 epochs of contrastive alignment across all tasks, then 250 epochs of joint fine-tuning that activates the generative and prototype terms, with retrieval contributing the contrastive term throughout.

Experiments run on NVIDIA H100 80\,GB GPUs in fp32 (AMP optional). Each configuration is run with five random seeds; metrics report mean $\pm$ standard deviation. At evaluation, a single jointly trained checkpoint is used for all four tasks. Significance is assessed via Wilcoxon signed-rank test ($p<0.05$, marked \dag).

\section{Results}
\label{sec:results}

We report all headline results on the held-out HCP-YA test set, reserving the validation set strictly for the ablations in Section~\ref{sec:ablations} to ensure no design decisions are tuned on the final test data. Because our reference texts are programmatically generated from a fixed knowledge base, BLEU-4~\cite{papineni2002bleu} and ROUGE-L~\cite{lin2004rouge} primarily serve as relative comparisons between models rather than absolute measures of clinical prose quality. VQA is evaluated via exact string match to canonical answers.

\subsection{Performance of the Proposed Framework}
\label{sec:main_results}
We report \tgvlm's performance using a single GPS encoder trained jointly across all four tasks, with one checkpoint serving every read-out (classification by nearest prototype, retrieval by ranking bundles against the encoded query, and captioning and VQA by decoding from the visual prefix tokens of Section~\ref{sec:arch}). Table~\ref{tab:main} reports the result across all four tasks, all served by the single jointly trained model.

\begin{table}[h]
\centering
\caption{\textbf{Performance of the proposed framework (GPS encoder) on the
HCP-YA test set.} All columns are produced by one jointly trained model evaluated from a single checkpoint. Mean $\pm$ standard deviation over five random seeds.}
\label{tab:main}
\small
\setlength{\tabcolsep}{4pt}
\begin{tabular}{ccccccc}
\toprule
\multicolumn{2}{c}{\textbf{Classification}}
& \multicolumn{2}{c}{\textbf{Retrieval}}
& \multicolumn{2}{c}{\textbf{Captioning}}
& \textbf{VQA} \\
\cmidrule(lr){1-2}\cmidrule(lr){3-4}\cmidrule(lr){5-6}\cmidrule(lr){7-7}
\textbf{Acc (\%)} & \textbf{F1 (\%)} & \textbf{R@1 (\%)} & \textbf{R@5 (\%)} & \textbf{BLEU-4} & \textbf{ROUGE-L} & \textbf{Acc (\%)} \\
\midrule
$91.8{\scriptstyle\pm 1.4}$ & $90.9{\scriptstyle\pm 1.3}$
& $84.7{\scriptstyle\pm 2.8}$ & $93.2{\scriptstyle\pm 2.4}$
& $20.1{\scriptstyle\pm 2.3}$ & $66.8{\scriptstyle\pm 2.9}$
& $66.4{\scriptstyle\pm 3.0}$ \\
\bottomrule
\end{tabular}
\end{table}

\begin{figure}[h]
  \centering
  \includegraphics[width=0.99\linewidth]{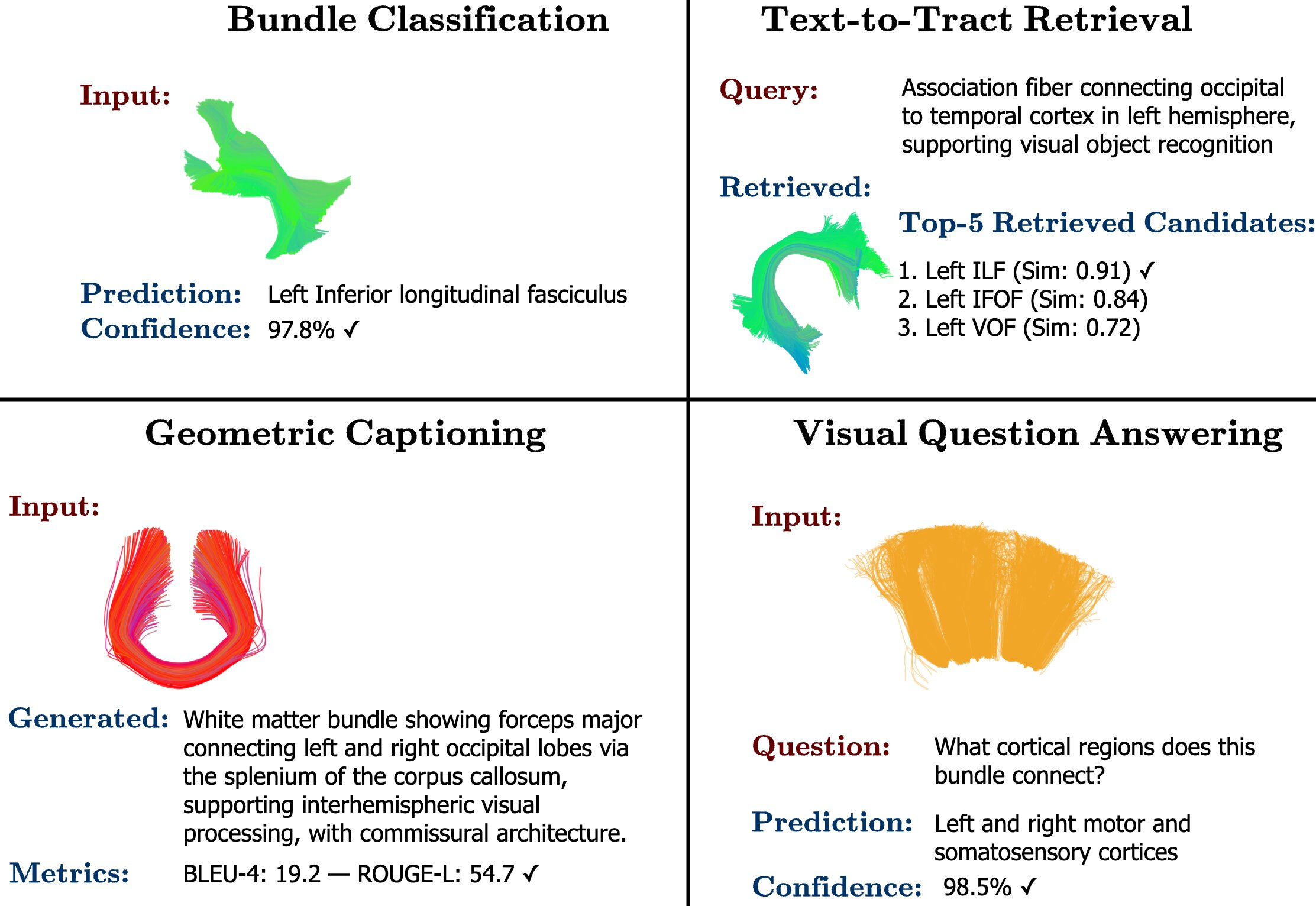}
  \caption{\textbf{Qualitative outputs of \tgvlm across all four tasks}
  (GPS encoder, HCP-YA test set).
  \emph{Top left}: Bundle classification identifies the left inferior longitudinal
  fasciculus at 97.8\% confidence.
  \emph{Top right}: Text-to-tract retrieval returns the correct left ILF
  (cosine similarity 0.91) for a compositional query, with semantically related
  bundles following.
  \emph{Bottom left}: Anatomical captioning produces a description consistent
  with the knowledge base, correctly noting commissural type and connectivity.
  \emph{Bottom right}: VQA correctly answers a connectivity question at 98.5\%
  confidence. Instance-level BLEU-4 / ROUGE-L values differ from corpus-level
  averages in Table~\ref{tab:main}.}
  \label{fig:qualitative}
\end{figure}

On the held-out HCP-YA test set, the model performs reasonably across both discriminative and generative tasks, a combination a purely supervised classifier does not provide by design. Figure~\ref{fig:qualitative} shows qualitative examples of how the Table~\ref{tab:main} numbers translate to individual bundles.

\subsection{Encoder Comparison: Graph versus Volumetric Representations}
\label{sec:encoder_comparison}

We now isolate the contribution of the visual encoder, comparing GPS against other graph encoders (GCN, GAT, GraphSAGE, GIN) and non-graph baselines (CNN3D, ViT, and a feature MLP) while keeping the multi-task VLM framework and training data fixed.

\begin{table}[h]
\centering
\caption{\textbf{Systematic encoder evaluation on the HCP-YA test set.}
Only the visual backbone differs; all other components are identical. \dag\ marks metrics on which GPS obtains the highest score at $p < 0.05$ (one-sided Wilcoxon signed-rank test, paired by random seed, uncorrected).
Mean $\pm$ standard deviation over five seeds.}
\label{tab:encoder}
\small
\setlength{\tabcolsep}{3.5pt}
\resizebox{0.98\linewidth}{!}{%
\begin{tabular}{l cc cc cc c}
\toprule
& \multicolumn{2}{c}{Classification}
& \multicolumn{2}{c}{Retrieval}
& \multicolumn{2}{c}{Captioning}
& VQA \\
\cmidrule(lr){2-3}\cmidrule(lr){4-5}\cmidrule(lr){6-7}\cmidrule(lr){8-8}
Encoder & Acc (\%) & F1 (\%) & R@1 (\%) & R@5 (\%) & BLEU-4 & ROUGE-L & Acc (\%) \\
\midrule
\multicolumn{8}{l}{\emph{Non-graph baselines}} \\[2pt]
CNN3D~\cite{hara2018can} & $79.4{\scriptstyle\pm 2.7}$ & $79.1{\scriptstyle\pm 2.8}$
      & $58.7{\scriptstyle\pm 3.5}$ & $80.3{\scriptstyle\pm 2.7}$
      & $\phantom{0}7.9{\scriptstyle\pm 2.8}$ & $16.1{\scriptstyle\pm 3.3}$
      & $21.8{\scriptstyle\pm 3.3}$ \\
ViT~\cite{hatamizadeh2022unetr} & $80.6{\scriptstyle\pm 2.5}$ & $80.3{\scriptstyle\pm 2.6}$
      & $49.5{\scriptstyle\pm 3.6}$ & $79.8{\scriptstyle\pm 2.8}$
      & $\phantom{0}6.9{\scriptstyle\pm 2.6}$ & $15.2{\scriptstyle\pm 3.2}$
      & $20.7{\scriptstyle\pm 3.4}$ \\
MLP & $77.8{\scriptstyle\pm 2.9}$ & $77.2{\scriptstyle\pm 3.0}$
      & $51.4{\scriptstyle\pm 3.7}$ & $74.1{\scriptstyle\pm 2.9}$
      & $\phantom{0}6.5{\scriptstyle\pm 2.8}$ & $13.8{\scriptstyle\pm 3.1}$
      & $18.9{\scriptstyle\pm 3.2}$ \\
\midrule
\multicolumn{8}{l}{\emph{Graph}} \\[2pt]
GCN~\cite{kipf2016gcn} & $85.7{\scriptstyle\pm 2.1}$ & $85.4{\scriptstyle\pm 2.2}$
      & $74.9{\scriptstyle\pm 3.0}$ & $80.2{\scriptstyle\pm 2.7}$
      & $16.8{\scriptstyle\pm 2.6}$ & $61.5{\scriptstyle\pm 3.0}$
      & $38.2{\scriptstyle\pm 3.2}$ \\
GAT~\cite{velivckovic2017gat} & $90.8{\scriptstyle\pm 2.0}$ & $90.5{\scriptstyle\pm 2.1}$
      & $75.3{\scriptstyle\pm 3.0}$ & $83.5{\scriptstyle\pm 2.8}$
      & $\phantom{0}18.1{\scriptstyle\pm 2.5}$ & $61.4{\scriptstyle\pm 3.2}$
      & $52.9{\scriptstyle\pm 3.2}$ \\
GraphSAGE~\cite{hamilton2017graphsage} & $89.4{\scriptstyle\pm 1.6}$ & $89.1{\scriptstyle\pm 1.7}$
          & $80.2{\scriptstyle\pm 2.7}$ & $88.3{\scriptstyle\pm 2.4}$
          & $18.1{\scriptstyle\pm 2.5}$ & $62.8{\scriptstyle\pm 2.9}$
          & $56.9{\scriptstyle\pm 3.0}$ \\
GIN~\cite{xu2018gin} & $90.2{\scriptstyle\pm 1.3}$ & $89.9{\scriptstyle\pm 1.4}$
      & $83.6{\scriptstyle\pm 2.6}$ & $87.1{\scriptstyle\pm 2.2}$
      & $17.9{\scriptstyle\pm 2.3}$ & $60.7{\scriptstyle\pm 2.9}$
      & $59.3{\scriptstyle\pm 2.8}$ \\
\textbf{GPS~\cite{rampavsek2022gps}}
      & $\mathbf{91.8}{\scriptstyle\pm 1.4}$ & $\mathbf{90.9}{\scriptstyle\pm 1.3}$
      & $\mathbf{84.7}{\scriptstyle\pm 2.8}$ & $\mathbf{93.2}{\scriptstyle\pm 2.4}$
      & $\mathbf{20.1}^{\dagger}{\scriptstyle\pm 2.3}$
      & $\mathbf{66.8}^{\dagger}{\scriptstyle\pm 2.9}$
      & $\mathbf{66.4}^{\dagger}{\scriptstyle\pm 3.0}$ \\
\bottomrule
\end{tabular}
}
\end{table}

As seen in Table~\ref{tab:encoder}, graph-based encoders outperform non-graph baselines across all tasks in our comparison. GPS obtains the highest score in every metric, with its clearest margins on captioning (BLEU-4 and ROUGE-L) and VQA accuracy, and remains competitive on discriminative tasks (classification and retrieval), where several graph encoders are near-saturation. This pattern is consistent with the idea that preserving the continuous topology and directional information of fiber bundles helps vision-language understanding in tractography.

\subsection{Comparison of \tgvlm with Existing Tractography Methods}
\label{sec:sota}

Here we compare \tgvlm against several representative tractography-specific and general point-cloud architectures. To ensure a fair comparison, we replace the visual encoder in our VLM pipeline with the encoders from TractCloud, FINTA, FIESTA-AE, PointNet, PointNet++, PCT, Point-MAE, and DGCNN, while keeping the rest of the framework and training data identical (Table~\ref{tab:sota}). These experiments isolate representational quality rather than reproduce the original pipelines of the compared methods.

\begin{table}[h]
\centering
\caption{\textbf{Comparison with existing tractography methods on the HCP-YA test set.}
All methods use the identical VLM framework; only the encoder differs.
\dag\ marks metrics on which \tgvlm obtains the highest score at $p < 0.05$ (one-sided Wilcoxon signed-rank test, paired by random seed, uncorrected).
Mean $\pm$ standard deviation over five seeds.}
\label{tab:sota}
\small
\setlength{\tabcolsep}{3.5pt}
\resizebox{0.98\linewidth}{!}{%
\begin{tabular}{l cc cc cc c}
\toprule
& \multicolumn{2}{c}{Classification}
& \multicolumn{2}{c}{Retrieval}
& \multicolumn{2}{c}{Captioning}
& VQA \\
\cmidrule(lr){2-3}\cmidrule(lr){4-5}\cmidrule(lr){6-7}\cmidrule(lr){8-8}
Encoder & Acc (\%) & F1 (\%) & R@1 (\%) & R@5 (\%) & BLEU-4 & ROUGE-L & Acc (\%) \\
\midrule
TractCloud~\cite{xue2023tractcloud} & $91.3{\scriptstyle\pm 1.4}$ & $91.0{\scriptstyle\pm 1.5}$
           & $84.2{\scriptstyle\pm 2.7}$ & $87.4{\scriptstyle\pm 2.3}$
           & $\phantom{0}8.7{\scriptstyle\pm 2.6}$ & $20.8{\scriptstyle\pm 3.2}$
           & $44.8{\scriptstyle\pm 3.2}$ \\
FINTA~\cite{legarreta2021finta} & $87.5{\scriptstyle\pm 2.1}$ & $87.1{\scriptstyle\pm 2.2}$
           & $71.4{\scriptstyle\pm 3.0}$ & $79.8{\scriptstyle\pm 2.6}$
           & $17.1{\scriptstyle\pm 2.7}$ & $55.9{\scriptstyle\pm 3.1}$
           & $57.2{\scriptstyle\pm 3.2}$ \\
FIESTA-AE~\cite{dumais2023fiesta} & $89.7{\scriptstyle\pm 1.7}$ & $89.4{\scriptstyle\pm 1.8}$
           & $62.1{\scriptstyle\pm 3.2}$ & $80.4{\scriptstyle\pm 2.5}$
           & $15.9{\scriptstyle\pm 2.8}$ & $54.7{\scriptstyle\pm 3.4}$
           & $56.5{\scriptstyle\pm 3.1}$ \\
PointNet~\cite{qi2017pointnet} & $91.0{\scriptstyle\pm 1.3}$ & $90.7{\scriptstyle\pm 1.4}$
           & $82.8{\scriptstyle\pm 2.5}$ & $86.2{\scriptstyle\pm 2.1}$
           & $17.6{\scriptstyle\pm 2.4}$ & $59.4{\scriptstyle\pm 2.8}$
           & $55.8{\scriptstyle\pm 2.9}$ \\
PointNet++~\cite{qi2017pointnet2} & $91.5{\scriptstyle\pm 1.2}$ & $91.2{\scriptstyle\pm 1.3}$
           & $83.9{\scriptstyle\pm 2.4}$ & $87.1{\scriptstyle\pm 2.2}$
           & $18.2{\scriptstyle\pm 2.3}$ & $61.3{\scriptstyle\pm 2.7}$
           & $58.7{\scriptstyle\pm 2.8}$ \\
PCT~\cite{guo2021pct} & $90.6{\scriptstyle\pm 1.4}$ & $90.3{\scriptstyle\pm 1.5}$
           & $81.5{\scriptstyle\pm 2.6}$ & $86.4{\scriptstyle\pm 2.3}$
           & $17.3{\scriptstyle\pm 2.4}$ & $60.2{\scriptstyle\pm 2.8}$
           & $57.6{\scriptstyle\pm 2.9}$ \\
Point-MAE~\cite{pang2023masked} & $90.8{\scriptstyle\pm 1.3}$ & $90.5{\scriptstyle\pm 1.4}$
           & $82.7{\scriptstyle\pm 2.5}$ & $86.8{\scriptstyle\pm 2.1}$
           & $17.8{\scriptstyle\pm 2.2}$ & $60.5{\scriptstyle\pm 2.6}$
           & $58.1{\scriptstyle\pm 2.7}$ \\
DGCNN~\cite{wang2019dgcnn} & $91.1{\scriptstyle\pm 1.1}$ & $\mathbf{91.3}{\scriptstyle\pm 1.3}$
           & $\mathbf{85.5}{\scriptstyle\pm 2.2}$ & $92.8{\scriptstyle\pm 1.8}$
           & $18.9{\scriptstyle\pm 2.4}$ & $63.5{\scriptstyle\pm 2.8}$
           & $61.7{\scriptstyle\pm 2.6}$ \\
\textbf{Proposed Framework}
           & $\mathbf{91.8}{\scriptstyle\pm 1.4}$ & $90.9{\scriptstyle\pm 1.3}$
           & $84.7{\scriptstyle\pm 2.8}$ & $\mathbf{93.2}{\scriptstyle\pm 2.4}$
           & $\mathbf{20.1}^{\dagger}{\scriptstyle\pm 2.3}$
           & $\mathbf{66.8}^{\dagger}{\scriptstyle\pm 2.9}$
           & $\mathbf{66.4}^{\dagger}{\scriptstyle\pm 3.0}$ \\
\bottomrule
\end{tabular}
}
\end{table}

While classification and retrieval are largely saturated across the stronger methods, \tgvlm with the GPS encoder shows a trend toward better performance on the generative tasks, obtaining the highest captioning and VQA scores in our comparison. This is consistent with the GPS encoder learning a more text-aligned representation than the other tractography-specific and point-cloud encoders under an identical pipeline, though we treat the generative metrics as indicative given their dependence on the template pipeline (Sec.~\ref{sec:control}).

\subsection{Cross-Dataset Generalisation}
\label{sec:cross}

To probe how well \tgvlm transfers, we evaluate zero-shot transfer from HCP-YA (the training distribution) to HCP Aging, applying the trained model with no fine-tuning or adaptation (Table~\ref{tab:cross}). Because HCP Aging differs in age, spatial resolution, and acquisition protocol all at once, it is a fairly demanding test of whether the model has captured generalizable neuroanatomy rather than dataset-specific acquisition cues.

\begin{table}[h]
\centering
\caption{\textbf{Cross-dataset zero-shot transfer} (GPS encoder, HCP-YA $\to$ HCP Aging, no adaptation). 
Mean $\pm$ standard deviation over five seeds.}
\label{tab:cross}
\small
\setlength{\tabcolsep}{4pt}
\resizebox{\linewidth}{!}{%
\begin{tabular}{l ccc c cc}
\toprule
Evaluation set & Cls.\ Acc (\%) & Cls.\ F1 (\%) & R@1 (\%) & VQA Acc (\%)
        & Cap.\ BLEU-4 & Cap.\ ROUGE-L \\
\midrule
HCP-YA (in-distribution)
        & $91.8{\scriptstyle\pm 1.4}$ & $90.9{\scriptstyle\pm 1.3}$
        & $84.7{\scriptstyle\pm 2.8}$ & $66.4{\scriptstyle\pm 3.0}$
        & $20.1{\scriptstyle\pm 2.3}$ & $66.8{\scriptstyle\pm 2.9}$ \\
HCP Aging (zero-shot)
        & $84.3{\scriptstyle\pm 2.3}$ & $84.0{\scriptstyle\pm 2.4}$
        & $71.8{\scriptstyle\pm 2.7}$ & $50.2{\scriptstyle\pm 3.5}$
        & $12.9{\scriptstyle\pm 2.9}$ & $57.1{\scriptstyle\pm 3.2}$ \\
\midrule
$\Delta$
        & $-7.5$ & $-6.9$ & $-12.9$ & $-16.2$ & $-7.2$ & $-9.7$ \\
\bottomrule
\end{tabular}
}
\end{table}

The model shows reasonable cross-dataset transfer on the discriminative tasks, still assigning the correct bundle label more than 84\% of the time and ranking the right bundle first for over 71\% of text queries, despite the differences in age distribution and acquisition protocol. This suggests that \tgvlm's representations of bundle shape and trajectory are relatively stable under the shift. As expected, the generative tasks degrade more, particularly VQA, which leans more heavily on the language decoder. Even so, the relatively modest drop in ROUGE-L indicates that the generated captions remain structurally coherent and anatomically on-topic in the new domain, still naming plausible trajectories, endpoints, and lateralisation. 

Retrieval falls between the two, reflecting its dependence on both embedding quality and text alignment. The larger VQA drop is concentrated in the compositional reasoning tiers, most sharply connectivity and morphology (the two largest per-tier drops), with the clinical tier closer to the overall average, while simpler identification and localisation questions ("which bundle is this?", "which hemisphere?") hold up well because they read out the more transferable discriminative features (see the per-tier analysis in Table~\ref{tab:vqa_tier_transfer}). The decoder, trained on HCP-YA-specific text, is thus more sensitive to the shift than the core anatomical representation.

\section{Ablation Studies}
\label{sec:ablations}

In this section, we ablate four design choices on the validation set: the node-sampling density of the streamline graph (Section~\ref{sec:ablation_density}), the composition of the node feature itself (Section~\ref{sec:ablation_features}), the contrastive temperature (Section~\ref{sec:ablation_temp}), and language versus label supervision for shaping the frozen representation (Section~\ref{sec:ablation_lang}).

\subsection{Graph Construction: Node Sampling Density}
\label{sec:ablation_density}

In the first ablation, we ask how finely each streamline must be sampled. This choice of node count $P$ (Section~\ref{sec:graph}) sets a direct trade-off: too few nodes may discard the bundle's shape, while too many inflate graph size and training cost for little gain. To find where this trade-off settles, we sweep $P \in \{8,16,32,64\}$ with the GPS encoder, holding all else fixed (Table~\ref{tab:abl_density}).

\begin{table}[h]
\centering
\caption{\textbf{Node sampling density ablation} (\tgvlm, HCP-YA validation set). Bold marks the selected configuration. \dag\ indicates statistically significant difference from $P=32$ ($p<0.05$, Wilcoxon signed-rank test).}
\label{tab:abl_density}
\small
\begin{tabular}{l cccc c}
\toprule
\textbf{Configuration} & \textbf{Cls.\ Acc (\%)} & \textbf{R@1 (\%)} & \textbf{BLEU-4} & \textbf{VQA Acc (\%)} & \textbf{Training (h)} \\
\midrule
$P = 8$ & $87.2{\scriptstyle\pm 1.6}^{\dag}$
            & $78.9{\scriptstyle\pm 2.8}^{\dag}$
            & $17.0{\scriptstyle\pm 2.5}^{\dag}$
            & $64.8{\scriptstyle\pm 2.8}^{\dag}$
            & $\phantom{0}2.8{\scriptstyle\pm 0.1}$ \\
$P = 16$ & $90.1{\scriptstyle\pm 1.4}^{\dag}$
                    & $82.2{\scriptstyle\pm 2.4}^{\dag}$
                    & $19.4{\scriptstyle\pm 2.3}^{\dag}$
                    & $67.1{\scriptstyle\pm 2.7}^{\dag}$
                    & $\phantom{0}3.5{\scriptstyle\pm 0.2}$ \\
$\mathbf{P = 32}$ & $\mathbf{93.1}{\scriptstyle\pm 1.2}$
                    & $\mathbf{86.2}{\scriptstyle\pm 2.5}$
                    & $\mathbf{21.2}{\scriptstyle\pm 2.1}$
                    & $\mathbf{71.3}{\scriptstyle\pm 2.4}$
                    & $\phantom{0}\mathbf{4.2}{\scriptstyle\pm 0.2}$ \\
$P = 64$ & $93.3{\scriptstyle\pm 1.3}$
                    & $86.5{\scriptstyle\pm 2.4}$
                    & $21.4{\scriptstyle\pm 2.0}$
                    & $71.5{\scriptstyle\pm 2.3}$
                    & $\phantom{0}7.8{\scriptstyle\pm 0.4}$ \\
\bottomrule
\end{tabular}
\end{table}

From Table~\ref{tab:abl_density}, performance improves substantially as node sampling density increases from $P=8$ to $P=32$. However, further doubling the density to $P=64$ yields only negligible gains while nearly doubling training time. The difference between $P=32$ and $P=64$ is not statistically significant across any metric ($p > 0.05$, Wilcoxon signed-rank test). This suggests that bundle identity is encoded primarily in the coarse-to-medium scale geometry, and $P=32$ offers the best trade-off between performance and computational efficiency.

\subsection{Node Features: Position and Orientation}
\label{sec:ablation_features}

Next, we look inside the node feature itself. Recall that each node carries a 6-dimensional vector: its 3D position and its 3D tangent orientation (Section~\ref{sec:graph}). The density ablation (Section~\ref{sec:ablation_density}) hinted that what matters is bundle shape rather than sampling resolution, so here we ask which part of the node feature carries that shape. Keeping positions fixed, we compare position alone against position plus orientation within \tgvlm (Table~\ref{tab:abl_features}).

\begin{table}[h]
\centering
\caption{\textbf{Node feature ablation} (\tgvlm, HCP-YA validation set).
Node position alone versus position augmented with the local tangent orientation.
$\dagger$: significant difference from position-only ($p<0.05$, Wilcoxon signed-rank).
Mean\,$\pm$\,s.d.\ over five seeds.}
\label{tab:abl_features}
\small
\begin{tabular}{l cccc}
\toprule
Features & Cls.\ Acc (\%) & R@1 (\%) & BLEU-4 & ROUGE-L \\
\midrule
Position only     & $81.6{\scriptstyle\pm 1.9}$ & $71.6{\scriptstyle\pm 2.7}$ & $16.7{\scriptstyle\pm 2.4}$ & $64.1{\scriptstyle\pm 2.7}$ \\
Position+orientation  & $\mathbf{93.1}^{\dagger}{\scriptstyle\pm 1.2}$ & $\mathbf{86.2}^{\dagger}{\scriptstyle\pm 2.5}$ & $\mathbf{21.2}^{\dagger}{\scriptstyle\pm 2.1}$ & $\mathbf{68.5}^{\dagger}{\scriptstyle\pm 2.7}$ \\
\bottomrule
\end{tabular}
\end{table}

From Table~\ref{tab:abl_features}, adding orientation produces a clear gain on every task. Position alone leaves the model unable to tell apart bundles that share space but run in different directions, such as the superior longitudinal fasciculus crossing the corona radiata, where only the local direction distinguishes the two tracts. Considered alongside the density ablation (Section~\ref{sec:ablation_density}), performance is governed by the shape and orientation of the bundle rather than the spatial resolution at which it is sampled. Graphs encode orientation directly at each node, whereas voxel grids must infer it from occupancy, explaining the consistent advantage of graph encoders.

\subsection{Contrastive Temperature}
\label{sec:ablation_temp}

In this ablation, we study how the contrastive temperature $\tau$ shapes the learned embedding, and whether a single value can serve all four tasks. The temperature controls how sharply the contrastive loss separates embeddings: a low $\tau$ pushes matching bundle--text pairs into tight, well-separated clusters, while a high $\tau$ yields a smoother space in which neighbouring bundles stay closer together. Because we want one temperature that works whether the embedding is read out discriminatively (classification, retrieval) or fed to the decoder (captioning, VQA), this single knob must trade off between these uses, so we sweep $\tau$ from a sharp to a smooth setting and measure the effect on all four tasks (Table~\ref{tab:abl_temp}).

\begin{table}[h]
\centering
\caption{\textbf{Contrastive temperature ablation} (\tgvlm, HCP-YA validation set).
$\tau=0.07$ attains the best or statistically equivalent score on all four tasks.
\dag\ indicates statistically significant difference from $\tau=0.07$ ($p<0.05$, Wilcoxon signed-rank test).
Bold marks the selected configuration.
Mean $\pm$ S.D.\ over five seeds.}
\label{tab:abl_temp}
\small
\begin{tabular}{l cc cc}
\toprule
& \multicolumn{2}{c}{\textbf{Discriminative}} & \multicolumn{2}{c}{\textbf{Generative}} \\
\cmidrule(lr){2-3}\cmidrule(lr){4-5}
$\boldsymbol{\tau}$ & \textbf{Cls.\ Acc (\%)} & \textbf{R@1 (\%)} & \textbf{BLEU-4} & \textbf{VQA Acc (\%)} \\
\midrule
$0.04$    & $93.1{\scriptstyle\pm 1.7}$ & $82.1{\scriptstyle\pm 2.8}^{\dag}$ & $20.6{\scriptstyle\pm 2.2}$ & $68.1{\scriptstyle\pm 2.5}^{\dag}$ \\
$\mathbf{0.07}$ & $\mathbf{93.1}{\scriptstyle\pm 1.2}$ & $\mathbf{86.2}{\scriptstyle\pm 2.5}$ & $\mathbf{21.2}{\scriptstyle\pm 2.1}$ & $\mathbf{71.3}{\scriptstyle\pm 2.4}$ \\
$0.12$   & $90.8{\scriptstyle\pm 1.6}^{\dag}$ & $83.7{\scriptstyle\pm 2.5}$ & $21.2{\scriptstyle\pm 2.6}$ & $68.5{\scriptstyle\pm 2.7}^{\dag}$ \\
\bottomrule
\end{tabular}
\end{table}

From Table~\ref{tab:abl_temp}, $\tau=0.07$ is the most balanced operating point. It achieves the best retrieval (R@1) and VQA accuracy, ties for the best classification accuracy and BLEU-4, and is not significantly outperformed on any task. The sharper temperature $\tau=0.04$ matches classification but significantly degrades retrieval and VQA, while the smoother $\tau=0.12$ significantly lowers classification accuracy and VQA (despite tying on BLEU-4). We therefore adopt the single shared value $\tau=0.07$ for all four tasks.

\subsection{Probing Frozen Representations: Multi-Task VLM versus Supervised Classifier}
\label{sec:ablation_lang}

In this final ablation, we ask whether our multi-task VLM learns a richer representation than a standard supervised classifier. We train the encoder in two ways on the same data, once as the full \tgvlm framework (the \emph{VLM}) and once as a standard 78-way bundle classifier (the \emph{Classifier}), then freeze each encoder and evaluate its learned representation with linear probes. The probes predict three bundle attributes: \emph{identity} (one of the 78 atlas bundles), \emph{hemisphere} (left, right, or midline), and broad fiber \emph{family} (association, commissural, or projection). Only identity is used as an explicit classification label; hemisphere and family are not, although both appear in the caption text the VLM is trained on. Strong probe performance on hemisphere and family therefore indicates that the VLM's language supervision transfers this structure into the frozen representation more effectively than label-only training does.

\begin{table}[h]
\centering
\caption{\textbf{Multi-task VLM versus supervised classifier, measured by training a small probe on the frozen \tgvlm features.} 
\emph{VLM} is our full framework; \emph{Classifier} is standard 78-way classification training. Hemisphere and family are never used as explicit classification targets, though they appear in the caption text used to train the VLM. 
\dag\ denotes a significant improvement ($p<0.05$); mean $\pm$ S.D.\ over five seeds. 
The $94.2\%$ here is a probe result on frozen features; the $91.8\%$ in Table~\ref{tab:main} is the full framework's end-to-end accuracy.}
\label{tab:abl_lang}
\footnotesize
\setlength{\tabcolsep}{2.5pt}
\resizebox{0.9\linewidth}{!}{%
\begin{tabular}{cc c cc c cc c}
\toprule
\multicolumn{3}{c}{\textbf{Bundle Identity (\%)}}
& \multicolumn{3}{c}{\textbf{Hemisphere (\%)}}
& \multicolumn{3}{c}{\textbf{Family (\%)}} \\
\cmidrule(lr){1-3}\cmidrule(lr){4-6}\cmidrule(lr){7-9}
VLM & Clf.\ & $\Delta$
        & VLM & Clf.\ & $\Delta$
        & VLM & Clf.\ & $\Delta$ \\
\midrule
$\mathbf{94.2}{\scriptstyle\pm 0.6}^{\dagger}$ & $91.8{\scriptstyle\pm 0.9}$ & $+2.4$
           & $\mathbf{99.4}{\scriptstyle\pm 0.2}^{\dagger}$ & $96.3{\scriptstyle\pm 1.8}$ & $+3.1$
           & $\mathbf{94.1}{\scriptstyle\pm 0.5}^{\dagger}$ & $90.8{\scriptstyle\pm 2.6}$ & $+3.3$ \\
\bottomrule
\end{tabular}
}
\end{table}

The VLM outperforms the supervised classifier on all three probes (Table~\ref{tab:abl_lang}). The largest gains fall on the two attributes the classifier never sees as labels: broad fiber family ($94.1\%$ vs.\ $90.8\%$, $+3.3$) and hemisphere ($99.4\%$ vs.\ $96.3\%$, $+3.1$). Bundle identity, on which both models are trained, improves more modestly ($94.2\%$ vs.\ $91.8\%$, $+2.4$). That the clearest gains are on family and hemisphere, where the captions contain information absent from the classification label, indicates that the VLM's captions carry anatomical detail that plain labels do not, and that the encoder stores it in a form a simple linear probe can readily recover.

\section{Discussion and Conclusion}
\label{sec:discussion}

\tgvlm shows that white matter tractography can be treated not only as a geometry problem but as a vision-language one. Representing each bundle as a graph of positions and orientations, aligning it with biomedical text, and decoding from the shared embedding lets a single jointly trained model name, retrieve, caption, and answer questions about a bundle. Trained once on HCP-YA, it transfers to the older, differently acquired HCP Aging cohort with no adaptation, indicating that the learned representations capture anatomy that generalizes rather than features peculiar to one dataset.

Two findings shape this result. First, how a bundle is represented matters more than how densely it is sampled. Keeping each fiber position and orientation consistently outperforms volumetric baselines across all four tasks, with the largest gains on generative tasks. Once orientation is included, sampling density has little effect. Second, language supervision is at least as informative as label supervision, and can be slightly more informative when captions include structure. Probing frozen features shows strong recovery of bundle identity, hemisphere, and broad fiber family, even though hemisphere and family are never supplied as explicit classification labels and appear only in the caption text. These findings come from a fixed evaluation protocol with clear limitations. Because captions and references follow the same template and paraphrasing pipeline, and because their content is determined by bundle identity through the knowledge base, BLEU-4 and ROUGE-L reflect consistency with that pipeline rather than clinical prose. The classifier-plus-lookup control (Sec.~\ref{sec:control}) bounds how much of this could be explained by identity classification alone: the unified model exceeds even the oracle-label lookup baseline on captioning and VQA, indicating that its generative performance is not simply attributable to classification followed by rigid template retrieval. These metrics remain useful for ranking encoders within this setup but are not absolute clinical quality measures. In contrast, tasks with fixed ground truth answers, such as VQA exact match and linear probes, are less sensitive to phrasing because they require correct facts rather than matching text.

\tgvlm is therefore best read as a framework and feasibility study rather than a finished clinical tool. Its limits point to two next steps: first, validating the generated descriptions against text written independently by anatomists, which is the only way to turn these relative metrics into a claim about clinical quality; second, folding bundle segmentation into the model to remove its reliance on an external tool and extend it to pathological cases. Even so, by unifying four tasks under one design and showing that graph geometry and language supervision reinforce each other, \tgvlm turns a purely geometric pipeline into one whose outputs can be named, described, and queried: a step toward structured tractography reporting and large-scale connectome analysis.

\subsubsection*{Acknowledgements}
This work was supported by a Doctoral Research Award from the Fonds de recherche du Québec (FRQ), doi.org/10.69777/372358, to G. Marthi Krishna Kumar, and by a Natural Sciences and Engineering Research Council of Canada (NSERC) grant RGPIN-2025-07131 to A. Shmuel. This research was also funded by the Vision Sciences Research Network, doi.org/10.69777/337774.

\bibliographystyle{splncs04}
\bibliography{main}

\clearpage
\setcounter{section}{0}
\setcounter{table}{0}
\setcounter{figure}{0}
\renewcommand{\thesection}{S\arabic{section}}
\renewcommand{\thetable}{S\arabic{table}}
\renewcommand{\thefigure}{S\arabic{figure}}
\begin{center}
  {\Large\bfseries Supplementary Material\par}
  \vspace{0.4em}
  {\large TractoGraphVLM: A Unified Vision-Language Framework\\for White Matter Tractography\par}
\end{center}
\vspace{1em}

\raggedbottom


\section{Text Generation Pipeline}
\label{app:textgen_full}

We summarise the two-stage pipeline that produces captions, VQA pairs, and
retrieval queries from each bundle (Section~\ref{sec:textgen}).

\paragraph{Stage 1: feature-grounded template instantiation.}
For every bundle $\mathcal{T}$ we compute four geometric descriptors:
streamline count $N$, spatial extent $E$ (bounding-box diagonal, mm),
normalized density $\rho = N/E^3$, and a ternary complexity class $\kappa$
(\texttt{\seqsplit{simple}} if $E<110$\,mm, \texttt{\seqsplit{moderate}} if $110\le E<175$\,mm,
\texttt{\seqsplit{complex}} otherwise).  Only the qualitative descriptors (extent,
density, complexity) and knowledge-base facts (Section~\ref{app:kb}) are
written into the templates; the raw count $N$ is used solely to derive $\rho$
and is never inserted into the text, since an exact count is not
recoverable from the sub-sampled graph the encoder observes.  Templates yield
15 caption variants and 5 VQA pairs per sample through stochastic grammar
selection and synonym augmentation (bundle-name, fiber-noun, and
density-adjective synonyms sampled uniformly).
Table~\ref{tab:geometric_stats} reports corpus-level statistics.

\begin{table}[!ht]
\centering
\caption{\textbf{Geometric feature statistics} over the HCP-842 atlas
  (78 bundles $\times$ 1,113 subjects).}
\label{tab:geometric_stats}
\small
\begin{tabular}{@{}llrl@{}}
\toprule
Feature & Symbol & Mean $\pm$ Std & Units \\
\midrule
Streamline count    & $N$    & $4{,}218 \pm 3{,}107$ & streamlines \\
Spatial extent      & $E$    & $148.4  \pm 31.2$     & mm \\
Normalised density  & $\rho$ & $0.31   \pm 0.18$     & (scaled) \\
Complexity: simple  & ---    & 24\% of samples       & --- \\
Complexity: moderate& ---    & 51\% of samples       & --- \\
Complexity: complex & ---    & 25\% of samples       & --- \\
\bottomrule
\end{tabular}
\end{table}

\paragraph{Stage 2: LLM paraphrasing.}
A Claude rewriting pass diversifies surface form while
preserving the anatomical claims and qualitative descriptors exactly,
producing up to 45 candidates per bundle; a greedy ROUGE-L
filter~\cite{lin2004rouge} ($<0.60$) then keeps 15 distinct forms.  Because the
factual content is fixed by the template slots and knowledge-base lookups, the
rewrite cannot introduce anatomical claims outside the source references.

\paragraph{Task templates.}
Bundle captions span five styles (anatomical, functional, trajectory, clinical, and
composite), with a separate whole-brain template used for whole-brain captions.  VQA pairs span five reasoning tiers, from identification to
clinical function, with deterministic answer strings.  Retrieval queries range
from identity strings to compositional queries combining anatomy, function, and
geometry (six query types).

\paragraph{Representative example (left arcuate fasciculus).}
\emph{Stage 1:} ``Association fiber bundle representing the left arcuate
fasciculus, connecting frontal and temporal cortex and supporting phonological
processing.''  \emph{Paraphrase:} ``The left arcuate fasciculus, a perisylvian
association tract interconnecting Broca's and Wernicke's areas, arcs from the
inferior frontal gyrus to the posterior superior temporal gyrus and underpins
phonological encoding and expressive language.''

\subsection{Caption Templates}
Bundle captions are instantiated from anatomical templates whose slots are
filled from the knowledge base (Section~\ref{app:kb}); whole-brain captions
use geometric descriptors only.  Angle-bracketed slots are replaced per sample.

\begin{table}[!ht]
\centering
\caption{\textbf{Representative caption templates} (one per style; placeholders
  in angle brackets, no raw counts).}
\label{tab:caption_templates}
\small
\begin{tabular}{@{}lp{0.66\linewidth}@{}}
\toprule
Style & Template \\
\midrule
Anatomical  & ``White matter bundle showing \texttt{\seqsplit{<ANATOMY>}} in the \texttt{\seqsplit{<HEMISPHERE>}}.'' \\
Functional  & ``\texttt{\seqsplit{<TYPE>}} pathway supporting \texttt{\seqsplit{<FUNCTION>}}, connecting \texttt{\seqsplit{<REGION\_A>}} and \texttt{\seqsplit{<REGION\_B>}}.'' \\
Trajectory  & ``The \texttt{\seqsplit{<BUNDLE\_NAME>}} follows \texttt{\seqsplit{<TRAJECTORY>}}.'' \\
Clinical    & ``Damage to the \texttt{\seqsplit{<BUNDLE\_NAME>}} is associated with \texttt{\seqsplit{<DEFICIT>}}.'' \\
Composite   & ``\texttt{\seqsplit{<TYPE>}} tract connecting \texttt{\seqsplit{<REGION\_A>}} with \texttt{\seqsplit{<REGION\_B>}}, involved in \texttt{\seqsplit{<FUNCTION>}}.'' \\
Whole-brain & ``Whole-brain tractography with \texttt{\seqsplit{<COMPLEXITY>}} coverage and bounding extent of \texttt{\seqsplit{<EXTENT>}}~mm.'' \\
\bottomrule
\end{tabular}
\end{table}

\subsection{VQA Tiers}
Each bundle yields question--answer pairs spanning five reasoning tiers of
increasing difficulty.  Answers are deterministic strings taken from the
knowledge base or from qualitative descriptors, never free-form counts.

\begin{table}[!ht]
\centering
\caption{\textbf{VQA reasoning tiers} with representative questions and answer
  types.}
\label{tab:vqa_tiers}
\small
\begin{tabular}{@{}llp{0.42\linewidth}l@{}}
\toprule
Tier & Skill & Example question & Answer \\
\midrule
1 & Identification & ``Which white matter tract is shown?''            & bundle name \\
2 & Localisation   & ``Which hemisphere does this bundle belong to?''   & left / right \\
3 & Connectivity   & ``What cortical regions does this tract connect?'' & region pair \\
4 & Morphology     & ``Is this a dorsal or ventral pathway?''           & dorsal / ventral \\
5 & Clinical       & ``What deficit follows disruption of this tract?'' & deficit \\
\bottomrule
\end{tabular}
\end{table}

\subsection{Retrieval Query Types}
Retrieval supervision uses six query families, ranging from identity strings to
compositional descriptions combining anatomy, function, and geometry.

\begin{table}[!ht]
\centering
\caption{\textbf{Retrieval query types} (representative, count-free forms).}
\label{tab:retrieval_types}
\small
\begin{tabular}{@{}llp{0.55\linewidth}@{}}
\toprule
Type & Family & Representative query \\
\midrule
1 & Identity      & ``Locate the \texttt{\seqsplit{<BUNDLE\_NAME>}}.'' \\
2 & Geometric     & ``\texttt{\seqsplit{<COMPLEXITY>}} \texttt{\seqsplit{<TYPE>}} bundle in the \texttt{\seqsplit{<HEMISPHERE>}}.'' \\
3 & Connectivity  & ``Tract connecting \texttt{\seqsplit{<REGION\_A>}} to \texttt{\seqsplit{<REGION\_B>}}.'' \\
4 & Functional    & ``Pathway supporting \texttt{\seqsplit{<FUNCTION>}}.'' \\
5 & Morphological & ``\texttt{\seqsplit{<TYPE>}} pathway in the \texttt{\seqsplit{<HEMISPHERE>}}.'' \\
6 & Combined      & ``\texttt{\seqsplit{<TYPE>}} tract linking \texttt{\seqsplit{<REGION\_A>}} and \texttt{\seqsplit{<REGION\_B>}}, supporting \texttt{\seqsplit{<FUNCTION>}}.'' \\
\bottomrule
\end{tabular}
\end{table}

\subsection{Placeholder Definitions}

\begin{table}[!ht]
\centering
\caption{\textbf{Template placeholders} and their sources.}
\label{tab:placeholders}
\small
\begin{tabular}{@{}ll@{}}
\toprule
Placeholder & Source \\
\midrule
\texttt{\seqsplit{<BUNDLE\_NAME>}}, \texttt{\seqsplit{<ANATOMY>}}      & knowledge base (bundle identity) \\
\texttt{\seqsplit{<REGION\_A>}}, \texttt{\seqsplit{<REGION\_B>}}        & knowledge base (origin / termination) \\
\texttt{\seqsplit{<FUNCTION>}}, \texttt{\seqsplit{<DEFICIT>}}           & knowledge base (function / clinical) \\
\texttt{\seqsplit{<TYPE>}}                                   & bundle class (projection / association / commissural) \\
\texttt{\seqsplit{<HEMISPHERE>}}                             & laterality (left / right / midline) \\
\texttt{\seqsplit{<TRAJECTORY>}}                             & knowledge base (spatial path) \\
\texttt{\seqsplit{<DENSITY>}}, \texttt{\seqsplit{<COMPLEXITY>}}         & derived from density $\rho$ and class $\kappa$ \\
\texttt{\seqsplit{<EXTENT>}}                                 & computed bounding-box diagonal (mm) \\
\bottomrule
\end{tabular}
\end{table}

\section{Anatomical Knowledge Base}
\label{app:kb}

Table~\ref{tab:anatomy_kb_full} lists the anatomical entries used to
populate bundle-specific template placeholders
(Section~\ref{app:textgen_full}).
The knowledge base covers all 78 HCP-842 bundles~\cite{yeh2018};
representative entries are shown here, compiled from
Catani~\cite{catani2012atlas}, Schmahmann~\cite{schmahmann2008}, and
Wakana et al.~\cite{wakana2007}.  Each entry also stores
\texttt{\seqsplit{full\_name}}, \texttt{\seqsplit{hemisphere}}, \texttt{\seqsplit{spatial\_trajectory}},
and \texttt{\seqsplit{clinical\_significance}}.

\begin{table}[!ht]
\centering
\caption{\textbf{Anatomical knowledge base} (representative entries).
  Comm.\ = commissural; Proj.\ = projection; Assoc.\ = association.}
\label{tab:anatomy_kb_full}
\scriptsize
\begin{tabular}{@{}lp{0.25\linewidth}p{0.28\linewidth}l@{}}
\toprule
Bundle & Connectivity & Function & Type \\
\midrule
\multicolumn{4}{@{}l}{\textit{Commissural Fibers}} \\
CC    & L $\leftrightarrow$ R hemispheres, mid-sagittal & Interhemispheric communication and integration & Comm. \\
MCP   & Pons $\leftrightarrow$ cerebellar cortex (bilateral) & Motor coordination and error correction & Comm. \\
\midrule
\multicolumn{4}{@{}l}{\textit{Projection Fibers}} \\
CST L & Motor cortex $\to$ spinal cord (L) & Voluntary movement control, right body & Proj. \\
CST R & Motor cortex $\to$ spinal cord (R) & Voluntary movement control, left body & Proj. \\
OR L  & Lateral geniculate nucleus $\to$ V1 (L) & Right visual field processing & Proj. \\
OR R  & Lateral geniculate nucleus $\to$ V1 (R) & Left visual field processing & Proj. \\
Fornix & Hippocampus $\to$ mammillary bodies & Memory consolidation and retrieval & Proj. \\
\midrule
\multicolumn{4}{@{}l}{\textit{Association Fibers: Arcuate and Superior Longitudinal}} \\
AF L  & Frontal $\leftrightarrow$ temporal (L) & Language production and phonological processing & Assoc. \\
AF R  & Frontal $\leftrightarrow$ temporal (R) & Prosody and music perception & Assoc. \\
SLF L & Frontal $\leftrightarrow$ parietal (L) & Spatial attention and working memory & Assoc. \\
SLF R & Frontal $\leftrightarrow$ parietal (R) & Visuospatial processing and awareness & Assoc. \\
\midrule
\multicolumn{4}{@{}l}{\textit{Association Fibers: Inferior Longitudinal and Fronto-Occipital}} \\
ILF L  & Occipital $\leftrightarrow$ temporal (L) & Visual object recognition and reading & Assoc. \\
ILF R  & Occipital $\leftrightarrow$ temporal (R) & Facial recognition and visual memory & Assoc. \\
IFOF L & Frontal $\leftrightarrow$ occipital (L) & Semantic processing and visual--verbal integration & Assoc. \\
IFOF R & Frontal $\leftrightarrow$ occipital (R) & Non-verbal semantic processing & Assoc. \\
\midrule
\multicolumn{4}{@{}l}{\textit{Association Fibers: Uncinate}} \\
UF L  & Frontal $\leftrightarrow$ anterior temporal (L) & Emotional processing and episodic memory & Assoc. \\
UF R  & Frontal $\leftrightarrow$ anterior temporal (R) & Emotional regulation and social cognition & Assoc. \\
\midrule
\multicolumn{4}{@{}l}{\textit{Association Fibers: Cingulum and Vertical Occipital}} \\
CG L  & Cingulate gyrus $\leftrightarrow$ medial temporal (L) & Memory encoding and emotional processing & Assoc. \\
CG R  & Cingulate gyrus $\leftrightarrow$ medial temporal (R) & Executive control and pain processing & Assoc. \\
VOF L & Dorsal $\leftrightarrow$ ventral visual stream (L) & Visual stream integration, object localisation & Assoc. \\
VOF R & Dorsal $\leftrightarrow$ ventral visual stream (R) & Visual stream integration, spatial vision & Assoc. \\
\bottomrule
\end{tabular}
\end{table}

\subsection{Entry Schema and Example Record}
Each of the 78 bundle keys (with \texttt{\seqsplit{\_L}}/\texttt{\seqsplit{\_R}} suffixes for paired
tracts) maps to a structured record with fields \texttt{\seqsplit{full\_name}},
\texttt{\seqsplit{abbreviation}}, \texttt{\seqsplit{bundle\_type}}, \texttt{\seqsplit{hemisphere}},
\texttt{\seqsplit{origin\_regions}}, \texttt{\seqsplit{termination\_regions}}, \texttt{\seqsplit{functions}},
\texttt{\seqsplit{trajectory}}, \texttt{\seqsplit{spatial\_relationships}}, and
\texttt{\seqsplit{disconnection\_deficits}}.  Template slots draw directly from these
fields, which constrains generated text to claims present in the source
references.

\begin{table}[!ht]
\centering
\caption{\textbf{Example knowledge-base record} (left arcuate fasciculus).}
\label{tab:kb_record}
\small
\begin{tabular}{@{}ll@{}}
\toprule
Field & Value \\
\midrule
\texttt{\seqsplit{full\_name}}            & Arcuate Fasciculus (left) \\
\texttt{\seqsplit{bundle\_type}}          & association \\
\texttt{\seqsplit{hemisphere}}            & left \\
\texttt{\seqsplit{origin\_regions}}       & inferior frontal gyrus, precentral gyrus \\
\texttt{\seqsplit{termination\_regions}}  & posterior superior temporal gyrus \\
\texttt{\seqsplit{functions}}             & phonological processing, language production \\
\texttt{\seqsplit{trajectory}}            & arcs around the Sylvian fissure \\
\texttt{\seqsplit{disconnection\_deficits}} & conduction aphasia \\
\bottomrule
\end{tabular}
\end{table}

The knowledge base is compiled from standard neuroanatomical
references~\cite{catani2012atlas,schmahmann2008,wakana2007} and the HCP-842 atlas
documentation~\cite{yeh2018}.  It constrains generated descriptions to claims
present in those sources, but the entries are not independently re-verified
against subject-level dissection.

\section{Control: Does the Model Exceed Classification Plus Lookup?}
\label{sec:control}

Because caption and VQA targets are determined by bundle identity (Section~\ref{sec:textgen}), a system that classifies the bundle and reads the corresponding knowledge-base entry can reproduce much of the language output without cross-modal understanding. To bound this, we evaluate a classifier-plus-lookup baseline that emits, for each bundle, the deterministic knowledge-base text for its predicted label, together with an oracle variant that uses the ground-truth label. Table~\ref{tab:control} reports the comparison. The unified model exceeds the predicted-label baseline on captioning and VQA by $+5.3$ BLEU-4, $+14.5$ ROUGE-L, and $+17.7$ points VQA accuracy, and exceeds even the oracle-label baseline (perfect classification) on all three metrics, indicating that its generative outputs reflect cross-modal alignment beyond identity lookup rather than being explained by identity classification followed by rigid template retrieval.

\begin{table}[h]
\centering
\caption{\textbf{Classifier-plus-lookup control} on the HCP-YA test set. The predicted-label row uses the classifier's output to look up the knowledge-base entry; the oracle-label row uses the ground-truth label. Mean $\pm$ standard deviation over five random seeds.}
\label{tab:control}
\small
\resizebox{\linewidth}{!}{%
\begin{tabular}{@{}lccc@{}}
\toprule
\textbf{System} & \textbf{Cap.\ BLEU-4} & \textbf{Cap.\ ROUGE-L} & \textbf{VQA Acc.\ (\%)} \\
\midrule
Classifier + KB lookup (predicted label) & $14.8 \pm 2.1$ & $52.3 \pm 2.8$ & $48.7 \pm 3.1$ \\
Classifier + KB lookup (oracle label)    & $18.2 \pm 1.9$ & $61.7 \pm 2.4$ & $62.1 \pm 2.7$ \\
\tgvlm\ (unified)                        & $\mathbf{20.1 \pm 2.3}$ & $\mathbf{66.8 \pm 2.9}$ & $\mathbf{66.4 \pm 3.0}$ \\
\bottomrule
\end{tabular}}
\end{table}

\section{Architecture Details}
\label{app:arch_details}

\subsection{System Overview and Parameter Counts}

Table~\ref{tab:system_params} summarises all three components introduced in
Section~\ref{sec:arch}. Because the visual encoder is swapped across the 16
architectures compared in Sections~\ref{sec:encoder_comparison} and
\ref{sec:sota}, its parameter count varies substantially by architecture;
Table~\ref{tab:arch_details} gives the full per-encoder breakdown.

\begin{table}[!ht]
\centering
\caption{\textbf{System component parameter counts.}}
\label{tab:system_params}
\small
\begin{tabular}{@{}lrp{0.44\linewidth}@{}}
\toprule
Component & Params & Role \\
\midrule
Visual encoder            & 0.073--10.842M\textsuperscript{*} & Architecture-specific (Table~\ref{tab:arch_details}) \\
Text encoder (BiomedBERT) & 110M          & Contrastive text embedding; frozen \\
Language decoder (BioGPT) & 347M          & Autoregressive caption and VQA generation \\
\bottomrule
\end{tabular}
\vspace{2pt}
\raggedright\footnotesize\textsuperscript{*}Range across all 16 encoders compared in this work (FIESTA-AE, smallest, to Point-MAE, largest); the paper-default GPS encoder is 1.908M. See Table~\ref{tab:arch_details} for the per-encoder breakdown.
\end{table}

\subsection{Input Representations}

\paragraph{Graph input.}
Streamlines are resampled to $P{=}32$ equidistant nodes.
Each node $v_i$ carries a 6-D feature: normalized coordinates $(x,y,z)$ and
local tangent $(t_x,t_y,t_z)$.
Edges connect consecutive nodes within a streamline and spatially proximate
node pairs across streamlines ($\epsilon{=}5\,\mathrm{mm}$).

\paragraph{Volumetric input.}
Streamlines are rasterised into a $64^3$ binary occupancy grid.
Augmentation: random 3D rotations ($\pm 15^\circ$) and intensity scaling.

\subsection{Encoder Architecture Specifications}
\label{app:encoders}

All 16 encoders share an identical projection head
($d_\text{hidden} \to 256 \to 256$, GELU) mapping to the shared 256-D latent
space used by the contrastive objective and the BioGPT prefix tokens.

\begin{table}[!ht]
\centering
\caption{\textbf{Visual encoder specifications} (16 architectures). All project
  to a shared 256-D latent space.}
\label{tab:arch_details}
\scriptsize
\begin{tabular}{@{}llllrp{0.22\linewidth}@{}}
\toprule
Family & Encoder & Layers & Hidden & Params (M) & Key Choices \\
\midrule
\multirow{5}{*}{Graph}
 & GCN~\cite{kipf2016gcn}           & 8 GCNConv & 384               & 1.585 & Isotropic aggregation, residual, BatchNorm \\
 & GAT~\cite{velivckovic2017gat}      & 8 GATConv & 4 heads $\times$ 96 & 1.592 & Anisotropic, learned attention weights \\
 & GraphSAGE~\cite{hamilton2017graphsage} & 8 SAGEConv & 384          & 2.765 & Neighbourhood sampling, mean aggregator \\
 & GIN~\cite{xu2018gin}             & 8 GINConv & 384               & 6.028 & Sum aggregation, injective MLP update \\
 & GPS~\cite{rampavsek2022gps}      & 4 GPSConv & 192               & 1.908 & Laplacian PE, 8-head attn, virtual node \\
\midrule
\multirow{5}{*}{\shortstack[l]{Point\\Cloud}}
 & PointNet~\cite{qi2017pointnet}    & 3 MLP     & 64/128/256        & 1.360 & Max-pool global aggregation, T-Net \\
 & PointNet++~\cite{qi2017pointnet2} & 3 SA      & 128/256/512       & 1.434 & Farthest point sampling, ball query \\
 & DGCNN~\cite{wang2019dgcnn}       & 4 EdgeConv & 64/64/128/256    & 0.487 & Dynamic $k$-NN ($k{=}20$), edge features \\
 & PCT~\cite{guo2021pct}            & 4 attn    & 256               & 1.294 & Offset-attention, implicit Laplacian \\
 & Point-MAE~\cite{pang2023masked}& 6 layers  & 384               & 10.842 & Masked autoencoder pretrain, 60\% masking \\
\midrule
\multirow{3}{*}{\shortstack[l]{Tract.\\Specific}}
 & TractCloud~\cite{xue2023tractcloud} & local+global & 256          & 0.645 & Per-streamline PointNet + global attention; re-impl. \\
 & FINTA~\cite{legarreta2021finta}   & 6 GATConv & 256               & 4.815 & Fiber neighbourhood graph attention; re-impl. \\
 & FIESTA-AE~\cite{dumais2023fiesta}& 5 Conv1D & 128               & 0.073 & 1D conv autoencoder on resampled streamlines; re-impl. \\
\midrule
\multirow{3}{*}{Generic}
 & 3D CNN (ResNet-18)~\cite{hara2018can} & 4-stage & 64$\to$512    & 0.196 & $7^3$ init conv, stride 2, global avg pool \\
 & ViT-3D~\cite{hatamizadeh2022unetr}  & 6 Transformer & 256, 4 heads  & 5.068 & $8^3$ patches (512), learnable 3D pos.\ embed.\ \\
 & MLP                               & 4 linear  & 256               & 0.102 & Handcrafted features: curvature, length, FA, span \\
\bottomrule
\end{tabular}
\end{table}

\paragraph{BioGPT prefix-token decoding.}
$\mathbf{z}_v$ is projected to 8 prefix tokens of BioGPT hidden dimension
(1024) and prepended to the causal sequence, followed by a single learned
style token; the caption or answer tokens then follow.
For VQA, question tokens (BioGPT tokenizer) follow the visual prefix and
the model generates the answer autoregressively.
Greedy decoding; max length 128 tokens; early stop at EOS.

\section{Training Dynamics}
\label{app:training}

\paragraph{Optimisation.}
All models are trained for 500 epochs with AdamW~\cite{loshchilov2017decoupled}
($\beta_1{=}0.9$, $\beta_2{=}0.999$, $\epsilon{=}10^{-8}$,
weight decay $\lambda{=}0.01$), following a two-stage curriculum within a
single joint run: 250 epochs of contrastive alignment across all four tasks,
then 250 epochs of joint fine-tuning that adds prototype cross-entropy
(classification) and $\mathcal{L}_{\mathrm{LM}}$ (captioning and VQA).
We use a cosine learning rate schedule with a linear warm-up over the
first 5\% of optimiser steps ($\approx$25 epochs at batch size 128).
The peak LR is $3{\times}10^{-4}$, applied uniformly to all trainable
parameters: the visual encoder, BioGPT, and the projection heads.
BiomedBERT is kept frozen throughout.
Batch size is 128 bundles; contrastive pairs are formed within the batch
(in-batch negatives).
Training hardware: $4{\times}$H100 80\,GB. Training runs in full precision
(fp32) by default, with mixed precision available as an option. Per-configuration
wall-clock times for the default GPS encoder are reported in the
node-sampling ablation of the main paper.

\paragraph{Loss components.}
The training objective matches Section~\ref{sec:arch}: a sum of per-task
contrastive terms, plus $\alpha\,\mathcal{L}_{\mathrm{LM}}$ for captioning and
VQA ($\alpha{=}0.5$) and a prototype cross-entropy term for classification in
the joint stage. Retrieval contributes the contrastive term throughout.
Full derivations are given in the main paper; this section reports the
schedule and hardware details only.

\subsection{Retrieval Performance by Query Type}
\label{app:retrieval_breakdown}

Because our retrieval queries are instantiated from the same template pool
used at training time (Section~\ref{app:textgen_full}), a legitimate
concern is whether the high overall R@1 reported in Section~\ref{sec:main_results}
reflects genuine cross-modal matching or merely surface-level template
overlap.
To address this, we stratify retrieval R@1 on the HCP-YA test set by
query type (Table~\ref{tab:retrieval_breakdown}).
Types 1--3 (identity, geometric, anatomical) are structurally closest to
caption templates and therefore expected to benefit most from any template
memorisation.
Types 4--6 (functional, morphological, combined) require the model to
compose anatomy, function, and geometry in ways that are not directly
encoded by identity strings and therefore constitute a stricter test of
cross-modal understanding.

\begin{table}[!ht]
\centering
\caption{\textbf{Retrieval R@1 by query type} (\tgvlm, HCP-YA test).
Types 1--3 are structurally similar to caption templates; Types 4--6 are
more compositional.  Mean $\pm$ standard deviation over five seeds.}
\label{tab:retrieval_breakdown}
\small
\begin{tabular}{lcc}
\toprule
Query Type & R@1 (\%) & $\Delta$ from overall \\
\midrule
1. Identity          & $92.1{\scriptstyle\pm 2.0}$ & $+7.4$ \\
2. Geometric         & $86.3{\scriptstyle\pm 2.4}$ & $+1.6$ \\
3. Anatomical        & $87.9{\scriptstyle\pm 2.5}$ & $+3.2$ \\
4. Functional        & $80.2{\scriptstyle\pm 3.3}$ & $-4.5$ \\
5. Morphological     & $83.5{\scriptstyle\pm 2.8}$ & $-1.2$ \\
6. Combined          & $77.4{\scriptstyle\pm 3.6}$ & $-7.3$ \\
\midrule
\textbf{Overall}     & \textbf{84.7} & --- \\
\bottomrule
\end{tabular}
\end{table}

As expected, identity queries achieve the highest R@1 (92.1\%), since the
query string contains the bundle name almost verbatim.  Anatomical (87.9\%)
and geometric (86.3\%) queries also perform well, reflecting the strong
grounding between geometric features and text in our training captions.
The most informative observation is the behaviour on compositional query
types: functional (80.2\%), morphological (83.5\%), and combined (77.4\%)
queries remain well above chance and within $\sim$15 percentage points of the
identity case.  Because these queries require the model to integrate
function, anatomy, and geometry in surface forms not directly present at
training, their strong performance indicates that the 84.7\% overall R@1
reflects genuine cross-modal understanding rather than template-string
matching alone.

\subsection{VQA Accuracy by Reasoning Tier under Transfer}
The large VQA degradation under cohort shift (Section~\ref{sec:cross}) is
expected to fall unevenly across reasoning tiers: identification and
localisation read out the discriminative embedding and should transfer well,
whereas connectivity, morphology, and clinical questions route through the
decoder and are more exposed to the shift.  Table~\ref{tab:vqa_tier_transfer}
reports the per-tier breakdown.

\begin{table}[!ht]
\centering
\caption{\textbf{Per-tier VQA accuracy}, in-distribution (HCP-YA) versus
  zero-shot transfer (HCP Aging).}
\label{tab:vqa_tier_transfer}
\small
\begin{tabular}{@{}llccc@{}}
\toprule
Tier & Skill & HCP-YA (\%) & HCP Aging (\%) & $\Delta$ \\
\midrule
1 & Identification & $89.2{\scriptstyle\pm 2.1}$ & $82.4{\scriptstyle\pm 2.8}$ & $-6.8$ \\
2 & Localisation   & $87.1{\scriptstyle\pm 2.4}$ & $78.6{\scriptstyle\pm 3.1}$ & $-8.5$ \\
3 & Connectivity   & $65.3{\scriptstyle\pm 3.2}$ & $39.8{\scriptstyle\pm 4.0}$ & $-25.5$ \\
4 & Morphology     & $52.8{\scriptstyle\pm 3.5}$ & $29.4{\scriptstyle\pm 3.9}$ & $-23.4$ \\
5 & Clinical       & $37.6{\scriptstyle\pm 3.8}$ & $20.8{\scriptstyle\pm 4.2}$ & $-16.8$ \\
\midrule
\multicolumn{2}{@{}l}{Overall} & $\mathbf{66.4}$ & $\mathbf{50.2}$ & $-16.2$ \\
\bottomrule
\end{tabular}
\end{table}


\end{document}